\documentclass[twocolumn,superscriptaddress,showpacs,aps,amsmath,amssymb]{revtex4}
\usepackage{epsfig}
\usepackage{amsbsy}
\usepackage{amsmath}
\usepackage{amsfonts}
\usepackage{graphicx}
\usepackage{latexsym}

\begin{document}

\title{Real-time dynamics of single-electron pumpings}
\author{Chuan-Yu Lin}
\affiliation{Department of Physics and Center for Quantum
information Science, National Cheng Kung University, Tainan 70101,
Taiwan}
\author{Wei-Min Zhang}
\email{wzhang@mail.ncku.edu.tw} \affiliation{Department of Physics
and Center for Quantum information Science, National Cheng Kung
University, Tainan 70101, Taiwan}

\begin{abstract}
In this Letter, we study the real-ime dynamics of single-electron pumpings. 
We modulate the left tunneling barrier along with the dot level but  
slightly changing the right barrier to operate the single-electron pumping 
device at zero bias. We show the adiabatic to the non-adiabatic transition
in pumping operations. We also find that the relation 
$I=e\omega/2\pi$ is only valid in the low operating frequency adiabatic regime.
\end{abstract}

\date{June 20,~2012}


 \maketitle

Charge pumps and turnstiles are nano-fabrication devices utilizing
controllable external gate voltages to manipulate electron transfer.
These devices are supposed to have important applications in quantum
metrology and solid-state quantum computation. Charge pumping
operations have been experimentally demonstrated with various double
resonance tunneling structures, in particular, through two
phase-shifted tunneling barriers\cite{setp1,set4,set5,set6,pumps}
at the zero and non-zero bias voltages. These experiments are aiming
at the goals of a high current at high frequency with high accuracy
of the pumps\cite{pumps,accurracy}, in order to meet the standard
quantum metrology error rate ($<10^{-7}$). As a subsequent work of
Ref.~\cite{Prework},  we study in this letter 
the single-electron pumping by modulating time-dependent tunneling
barriers along with the energy level of the dot through the external
gate voltages at zero-bias.

By modulating the tunnel barriers and the
dot level, it has been observed\cite{setp1,set4,set5,set6,pumps}
that electrons can be transferred one-by-one between the source and
the drain. In our previous work\cite{Prework}, we have studied the
time-dependent single-electron dynamics by modulating the left and
right tunneling barriers in anti-phase but fixing the dot level,
with a finite bias voltage, to examine the high frequency limit of
pumping operations. Here we shall follow the experimental setup
given in Ref.~\cite{set4,set5,pumps} for a different operating mode,
namely modulating the left tunneling barrier along with the dot level
but fixing or slightly changing the right barrier, and operate the
device at zero bias. We should study both the adiabatic and
non-adiabatic pumpings through different operating frequencies.

Utilizing the recently developed non-equilibrium quantum transport
theory\cite{Jin10083013} which is derived from the exact master
equation for nanoelectronics\cite{Tu08235311}, we obtain the
time-dependent electron occupation in the dot and the time-dependent
electron current flowing from each lead into the dot:
\begin{subequations}
\label{curr-char}
\begin{align}
 &n(t) =  v(t,t)+ u(t,t_0)n(t_0)u^{\dag}(t,t_0), \label{cc-a} \\
 &I_{\alpha}(t)= -{2e\over \hbar}{\rm Re}\int_{t_0}^{t} d\tau {\rm
Tr}\Big\{g_{\alpha}(t,\tau)v(\tau,t)-\widetilde{g}_{\alpha}(t,\tau)\notag
\\ & ~~~~~~ \times u^\dag(t,\tau)
+g_{\alpha}(t,\tau)u(\tau,t_0)n(t_0)u^{\dag}(t, t_0)\Big\} .
\label{cc-b}
\end{align}
\end{subequations}
Here $\alpha=L,R$ denote the left and right leads (the source
and the drain), and $n(t_0)$ is the initial electron occupation in the
dot.  The function $u(\tau, t_0)$ and $v(\tau,t)$ are the two-point 
Green functions that satisfies the Dyson equation\cite{Tu08235311,Jin10083013},
and the non-local time-correlation function is given by
$g_{\alpha}( \tau ,\tau') =\int \frac{d\omega}{2\pi} J_{\alpha}\left(
\omega, \tau, \tau'\right) e^{-i\omega ( \tau -\tau')}$ and
$\widetilde{g}_{\alpha}(\tau ,\tau') =\int \frac{d\omega}{2\pi}
J_{\alpha}( \omega,\tau, \tau' ) f_{\alpha}(\omega)e^{-i\omega ( \tau
-\tau')} $
with $g (\tau ,\tau')=g_L (\tau ,\tau')+g_R (\tau ,\tau')$ and $
\widetilde{g} (\tau ,\tau')= \widetilde{g}_L (\tau
,\tau')+\widetilde{g}_R (\tau ,\tau')$, where $f_{\alpha}(\omega)
=\frac{1}{e^{\beta (\omega-\mu_{\alpha})} +1}$ is the initial electron
distribution function of lead $\alpha$ at the inverse initial
temperature $\beta=1/k_BT$, and $\mu_{\alpha}$ the corresponding
chemical potential. The spectral density $J_{\alpha}(\omega,\tau, \tau')=2\pi
\varrho_{\alpha}(\omega)V_{\alpha}(\tau)V^*_{\alpha}(\tau')$,
where $\varrho_{\alpha}(\omega)$ is the
density of states of the lead and $V_{\alpha}(\tau)$ the lead-dot
coupling coefficient. The integrate kernels $g( \tau ,\tau')$ and
$\widetilde{g}(\tau ,\tau')$ take into account all the non-Markovian
back-reaction memory effects from the leads (reservoirs) associating
with quantum dissipation and fluctuation. 

Similar to the previous work\cite{Prework}, we take the spectral
densities with a Lorentzian-type shape,
\begin{align}
J_{\alpha}(\omega, \tau, \tau') = \frac{V_{\alpha}(\tau)V^*_{\alpha}(\tau')
W_{\alpha}^{2}}{(\omega -\mu_{\alpha})^{2}+W_{\alpha}^{2}},\label{spectral}
\end{align}
where $V_{\alpha}(\tau)$ are the time-dependent electron tunneling
amplitude between lead $\alpha$ and the dot that can be controlled by
the gate voltage $V_{G\alpha}(t)$, and $W_{\alpha}$ is the bandwidth
of the spectral density. 
The explicit relation between the tunneling rate and the gate voltage can be
obtained by solving the Schr\"{o}dinger equation for a
one-dimensional scattering problem. Here we obtain
\begin{align}
V_{\alpha}(t) \simeq \frac{2}{\sqrt{A(t)\cosh^2(2k_{\alpha}(t)a)+B(t)\sinh(2k_{\alpha}(t)a)}},\label{rate}
\end{align}
where $A(t)= (1+\frac{k_2}{k_1})^2$ and
$B(t)=(\frac{k_1k_2-k^{2}_{\alpha}}{k_1k_{\alpha}})$ with
$k_{\alpha}(t)$
=$\sqrt{\frac{2m^*(eV_{G\alpha}(t)-\mu)}{\hbar^{2}}}$,
$k_1$=$\sqrt{\frac{2m^*\mu}{\hbar}}$ and
$k_2$=$\sqrt{\frac{2m^*\varepsilon(t)}{\hbar}}$, where $a$ is the
width of the barriers, and $m^*$ is the effective mass of the
electron in the sample. We apply different sinusoidal wave
modulation to the tunneling barriers and the dot level:
$V_{G\alpha}(t)=V_{\alpha}^{dc}-V_{\alpha}^{ac}\cos(\omega_c t)$,
$\varepsilon(t)=\varepsilon_0-\varepsilon_c\cos(\omega_c t)$.
It shows that the tunneling rate between lead $\alpha$ and the dot is determined
not only by the gate voltage $V_\alpha (t)$ acting on the barrier 
between them but also the gate voltage $\varepsilon(t)$ acting on 
the dot level.
\begin{figure}
\includegraphics[width=0.7\columnwidth,angle=0]{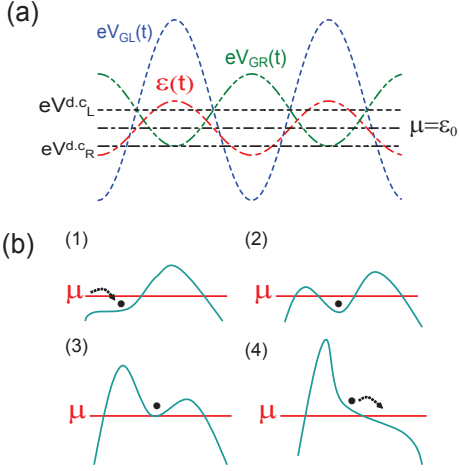}
\caption{(color online) (a) The driving fields on the left gate (blue dot
line), the right gate (green dash-dot-dot line), and the dot level
(red dash-dot line):
$V_{GL}(t)=V_{L}^{dc}-V_{L}^{ac}\cos(\omega_c t)$,
$V_{GR}(t)=V_{R}^{dc}+V_{R}^{ac}\cos(\omega_c t)$, and
$\varepsilon(t)=\varepsilon_0-\varepsilon_c\cos(\omega_c t)$. (b)
Schematic plots of the barriers changing at four different times
in each cycle.} \label{fig1}
\end{figure}

Figure \ref{fig1}(a) shows the corresponding driven fields:
$V_{GL}(t)$, $V_{GR}(t)$, and $\varepsilon(t)$, and their relative
phase. $V_{\alpha}^{dc}$ and $V_{\alpha}^{ac}$ are the external d.c
and a.c field applied on the barrier, $\varepsilon_{0}$ is the
energy of the dot level, and $\varepsilon_{c}$ is the oscillation
amplitude of the ac gate field applied on the dot.
Fig.~\ref{fig1}(b) show a schematic barrier
modulating process. In the first half of the pumping
operation, (1)-(2), the left barrier is opened to the source and the
right barrier is closed. The electron tunnel from the source
into the dot. In the second half of the cycle, (3)-(4), the electrons
transfers from the dot to the drain because the left barrier is
closed and the right barrier is opened.

\begin{figure}
\includegraphics[width=1.0\columnwidth,angle=0]{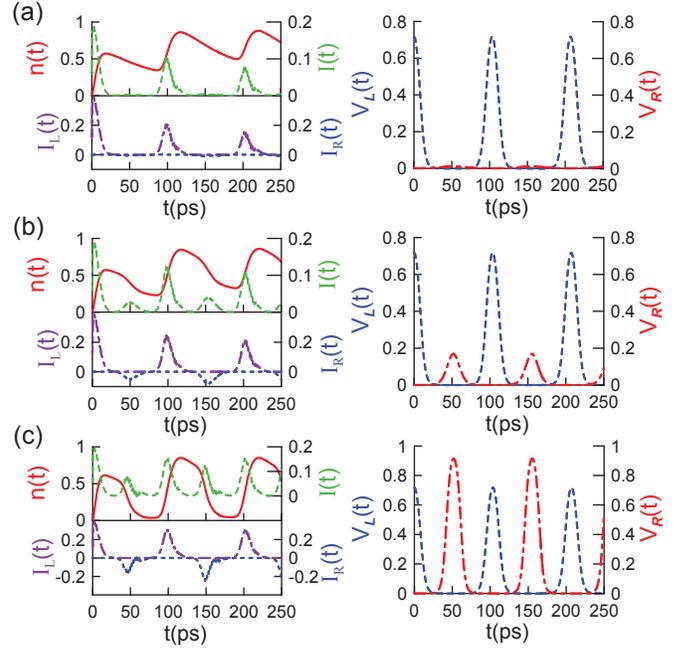}\newline
\caption{(color online) The left column shows the time-dependent electron
occupation (the red solid line), the pumping current (the green
dash-dash line), left current (purple dash-dot line) and right
current (blue dotted-dotted line) altered by the sinusoidal wave
modulation on the left tunneling barrier with $eV_{L}^{dc}
=1.6$meV and $eV_{L}^{ac}=0.8$ meV, along with the dot level with
$\varepsilon_c=0.2$ meV, but the right barrier with $eV_{R}^{dc}
=1.6$meV and (a) $eV_{R}^{ac}=0$, (b) $eV_{R}^{ac}=0.4$ meV, and (c)
$eV_{R}^{ac}=0.8$ meV. The driving field frequency $\omega_c=10$GHz.
The right column shows the corresponding time-dependent tunneling
amplitude of the left and right barriers. The initial temperature of
the source and drain is set at $kT \simeq 116mK$} \label{fig2}
\end{figure}

Now, we analyze how the electron turnstile operates with time-dependent 
tunneling barriers along with the dot level at zero bias. To be specific,
we fix the bandwidth of the spectral density by $W_{L,R}=2$ meV,
and the initial temperature of the leads at $k_BT=0.01$ meV, i.e. 
$T \simeq 116$mK. 
We begin with an equal dc voltage on both the left and the right
barriers, $eV_{L}^{dc}=eV_{R}^{dc}=1.6$ meV. Fig.~\ref{fig2} shows
the time-dependent electron population in the dot, the left and the right 
currents flowing into the dot as well as the pumping current
$I(t)=\frac{1}{2}[I_L(t)-I_R(t)]$ with a sinusoidal wave gate
voltage modulating the left tunneling barriers along with the dot
level.  The ac field strength on the left barrier and the dot level  
are given by $eV_{L}^{ac}=0.8$ meV and $\varepsilon_c=0.2$meV. All 
these parameters are adjustable in experiments. We find that with
zero ac field on the right barrier,  the tunneling 
amplitude for electrons transiting from the dot to the drain is almost zero 
[see the right plot of Fig.~\ref{fig2}(a)] to make electrons 
tunnel through the device. i.e. $I_R(t) \simeq 0$ as shown in the left plot of 
Fig.~\ref{fig2}(a). Increasing $eV_{R}^{ac}=0.4$ meV, the
corresponding tunneling amplitude is still not large enough to
make the single-electron pumping, see Fig.~\ref{fig2}(b). 
Only when the ac field strength is increased upon $eV_{R}^{ac}
=0.8$ meV, electrons can transit one by one, see Fig.~\ref{fig2}(c). 

\begin{figure}
\includegraphics[width=1.0\columnwidth,angle=0]{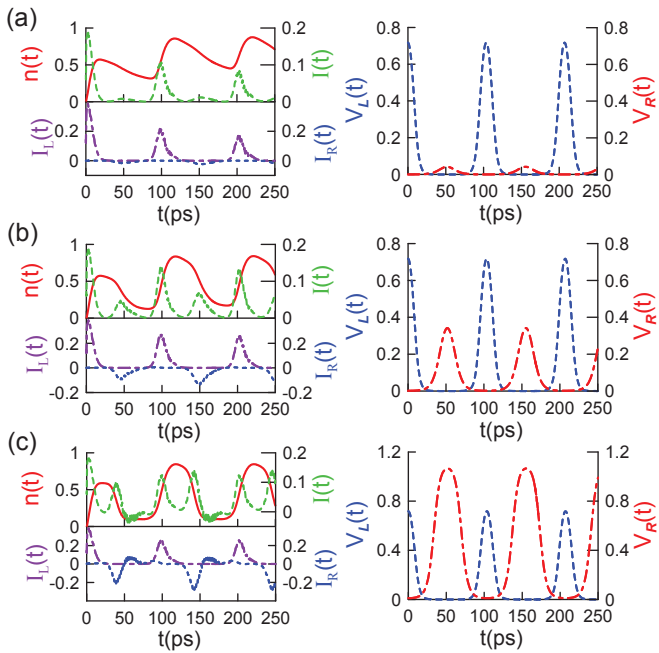}\newline
\caption{(color online) The same plots as Fig.~\ref{fig2} with different setup. 
Here the ac field $eV_{R}^{ac}=0.2$ meV is fixed but the dc field is changed to
(a) $eV_{R}^{dc}=1.6$ meV, (b) $eV_{R}^{dc}=1.3$ meV, and (c)
$eV_{R}^{dc}=1.0$ meV. The other parameters are the same as in Fig.~{\ref{fig2}}.}
\label{fig3}
\end{figure}

To make single-electron pumping operations with a fixed or small 
modulated right barrier, we may  check the dc voltage effect 
on the right barrier under a small ac field along with the dot level, e.g. 
$V_{R}^{ac}=\varepsilon_c=0.2$ meV. Figures \ref{fig3}(a) to (c) show that
a smaller dc field acting on the right barrier can increase the amplitude of the electron 
occupation oscillation. However, the overlap of the left and right currents
(see Fig.~\ref{fig3}(c)) indicates that electrons also flow back from the 
drain to the dot in the second half of the pumping cycle (as a leakage effect). 
In other words, reducing the dc field on the right barrier messes up the 
electrons pumping direction.

\begin{figure}
\includegraphics[width=1.0\columnwidth,angle=0]{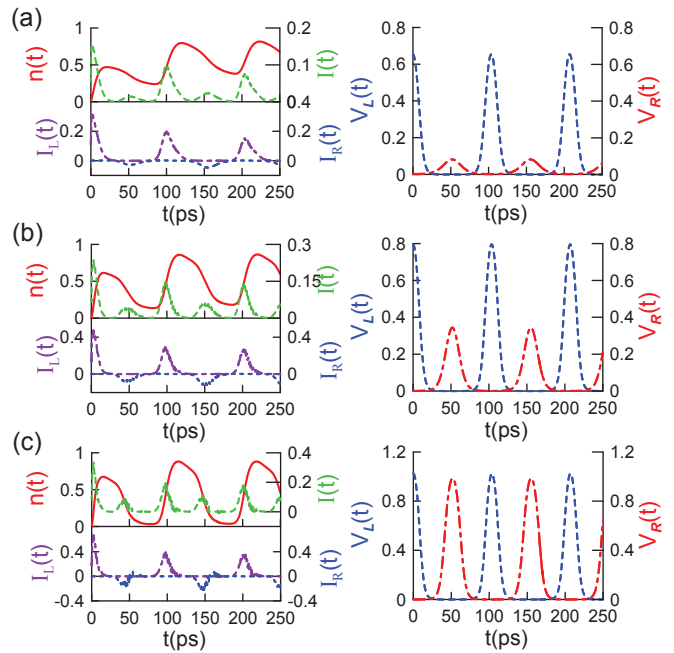}\newline
\caption{(color online) The same plots as Fig.~\ref{fig2} with fixing
$eV_{R}^{dc}=1.4$ meV and $eV_{R}^{ac}=0.2$meV but changing
the ac field on the dot level with (a) $\varepsilon_c=0.1$ meV,
(b) $\varepsilon_c=0.3$ meV, and (c) $\varepsilon_c=0.5$ meV. 
The other parameters are the same as in Fig.~{\ref{fig2}}}
\label{fig4}
\end{figure}

The remaining controllability is the ac field on the dot level. We take 
$V_{R}^{dc}=1.4$ meV so that the leakage effect is negligible, and keep 
the ac field on the right barrier relatively small, $eV_{R}^{ac}=0.2$ meV, 
compared to $eV_{R}^{ac}=0.8$ meV.  Figures~\ref{fig4} shows the result 
of electron transition with different ac field acting on the dot level. 
For a small ac field, $\varepsilon_c=0.1$ meV,  electrons cannot tunnel 
through the dot one by one because of the small tunneling amplitude, see Fig.~\ref{fig4}(a). 
Increasing $\varepsilon_c$ ($=0.3$ meV in Fig.~\ref{fig4}(b)) makes the 
tunneling amplitude larger, but not larger enough to reach the single-electron pumping. 
When $\varepsilon_c$ is increased to $0.5$ meV, the electron can tunnel 
through the dot one by one, as shown in Fig.~\ref{fig4}(c). 

Putting all the above results together shows that a good single-electron pumping 
operation, by modulating the left tunneling barrier along with the dot level
but fixing or slightly changing the right barrier at zero bias, requires 
(i) a high dc field on the left barrier with a slightly low dc field on the
right barrier to avoid electrons from leakage; and (ii)  a large
overlapping area between $\varepsilon(t)$ and $eV_{GR}(t)$ to ensure
electrons transferring through the dot one by one. A relatively large $\varepsilon(t)$
can actually induce a large tunneling amplitude between the dot and the right 
lead because of the off-phase with the small ac field on the right barrier. Thus the 
mechanism for single-electron pumping operations by modulating the left tunneling 
barrier along with the dot level but fixing or slightly changing the right 
barrier at zero bias is indeed equivalent to  the single-electron pumping 
through modulating the left and right tunneling barriers in anti-phase but 
fixing the dot level at a finite bias\cite{Prework}.

\begin{figure}
\includegraphics[width=0.9\columnwidth,angle=0]{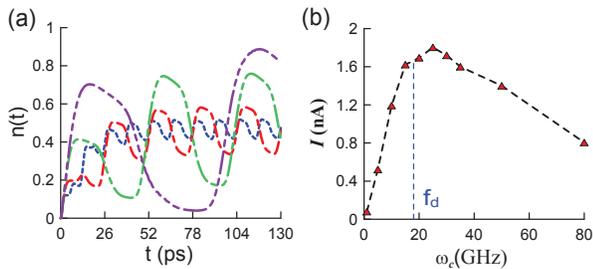}\newline
\caption{The adiabatic to the non-adiabatic transition
via the modulation frequency. (a) The electron occupation in the dot with
$f_{sin}=10$ GHz (purple dash-dotted-dotted-dotted line), $20$ GHz (green
dash-dotted-dotted line), $40$ GHz (red dash-dotted line),
and $80$ GHz (blue dash line). (b) The pumping current as a
function of the modulation frequency. Other parameters are the same
in Fig.~\ref{fig4}(c).} \label{fig5}
\end{figure}

Next we examine the single-electron pumping with different operation 
frequencies. As we have shown\cite{Prework}, there is a
character time, i.e. the dwell time or dwell frequency $\omega_d$. 
When the modulation frequency is smaller than the
dwell frequency $\omega_{c}<\omega_d$, the electron have enough time
to tunnel into the dot. Therefore, the electron have the higher
probability to be adiabatically transited one by one in each cycle. In the high
frequency regime, $\omega_{c}>\omega_d$, on the contrary, only
an electron can partially tunnel from the source to the drain (a non-adiabatic process).
Figure \ref{fig5}(a) shows the transition from adiabatic to non-adiabatic 
regime for the electron occupation in the dot. In the adiabatic regime, 
the variation of the electron occupation is almost one because the pumping 
period is much longer than the intrinsic time scale of the system. 
However, When the modulation frequency locates in the non-adiabatic regime, 
the electron cannot have enough time to response to the external driving field 
so that the electron occupation in the dot varies fractionally.
Moreover, Fig.~\ref{fig5}(b) gives the average current with respect to 
the driving frequency. It shows that for $\omega_{c}<\omega_d/2$, 
the pumping current has a linear relation with the modulation frequency,
$I=e\omega/2\pi$. When the operating frequency $\omega_{c}>\omega_d/2$, 
the frequency dependence of the transport current deviates from the linear 
regime substantially.  A similar result is obtained in
Ref.~\cite{nonadiabatic} recently.

In conclusion, the analysis based on the real-time electron transport shows
that the dc field strength, the amplitude of driving ac field and
the modulation frequency make the single-electron pumping device 
behave significantly different.  The larger overlapping area
between $\varepsilon(t)$ and $V_{GR}(t)$, and the relatively lower 
operating frequency are the two requirements for increasing the 
accuracy of single-electron pumpings and making the device a good
candidate for metrological applications.

This work is supported by the National Science Council of ROC under
Contract No. NSC-99-2112-M-006-008-MY3 and National Center for
Theoretical Science.

\end{document}